\documentclass[12pt]{article}
\usepackage{amsmath}
\usepackage{mathtools}
\usepackage{amssymb}
\usepackage{amsfonts}
\usepackage{graphicx}

\title{\vspace{-4.0cm} Implementation of a simultaneous message passing protocol using optical vortices}
\date{\vspace{-1.0cm} June, 2020}
\begin{document}
\maketitle
\begingroup  
\noindent Mateusz Szatkowski$^{1,2}$, Julian Koechlin$^{1,3}$, and Dorilian Lopez-Mago$^{1,*}$
\\
{\footnotesize 
$^{1}$ Tecnologico de Monterrey, Escuela de Ingenier\'{i}a y Ciencias, Ave. Eugenio Garza Sada 2501, Monterrey, N.L. 64849, Mexico.\\ \noindent
$^{2}$ Wroc\l{}aw University of Science and Technology, Department of Optics and Photonics, Wybrze\.{z}e Wyspia\'{n}skiego 27, 50-370 Wroc\l{}aw, Poland.\\ \noindent
$^{3}$ Department of Physics, University of Basel, 4056 Basel, Switzerland\\
$^\ast$ dlopezmago@tec.mx }
\endgroup

\section*{Abstract}
The implementation of optical quantum gates comes at the cost of incorporating a source of nonclassical light, which suffers from a low flux of photons, and thus, long acquisition times. Quantum-mimetic optical gates combine the benefits of quantum systems with the convenience of using intense light beams. Here, we are concerned with the classical implementation of a controlled-SWAP (c-SWAP) gate using the tools of structured light. A c-SWAP gate is a three-qubit gate, where one of the input qubits controls the exchange of information between the other two qubits. We use Laguerre-Gauss beams to realize the c-SWAP gate and demonstrate one of its primary applications: the comparison of two signals without revealing their content, i.e., a \textit{simultaneous message-passing} protocol. We achieve signal-comparison measurements with modulation speeds of kHz using a digital micromirror device as a spatial light modulator.  Our system is capable of performing dynamic error analysis and a normalization procedure, which overcomes the necessity of data postprocessing and largely reduces comparison times.

\section{Introduction}
In a simultaneous message-passing (SMP) protocol, two parties (e.g., Alice and Bob) send a message to a referee, who has the task of comparing both messages and outcomes their similarity, without exposing their content. The protocol has applications in quantum communications. For instance, Buhrman et al. used it to demonstrate quantum fingerprinting~\cite{Buhrman2001}, where Alice and Bob codify their messages in shorter strings called fingerprints by using quantum states~\cite{Arrazola2014, Xu2015}. In comparison with classical fingerprinting, the quantum variant reduces the amount of information needed to codify the message~\cite{Guan2016}. The improvement grows exponentially with the length of the original message.

\begin{figure}[htbp]
\centering
\includegraphics[width=10 cm]{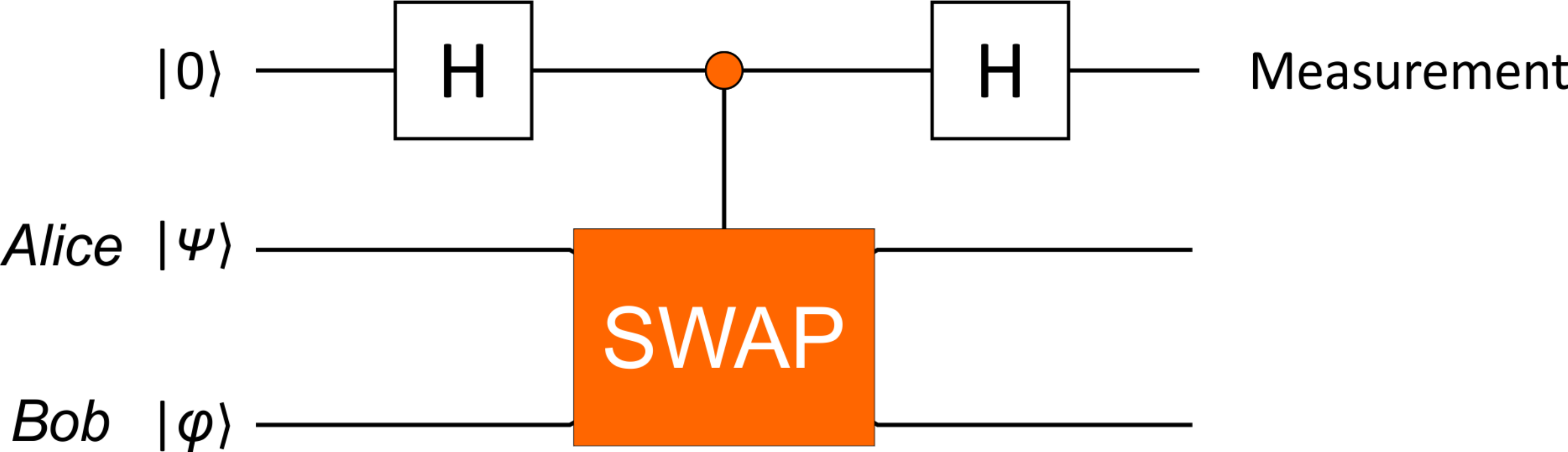}
\caption{SWAP test circuit to determine the similarity between the states $|\psi\rangle$ and $|\varphi\rangle$. H is the Hadamard gate.
 \label{fig:1}}
\end{figure}

Data exchange and data comparison are desirable operations to incorporate in an optical communication system. Recently, Urrego et al.~\cite{Urrego2020} experimentally demonstrate a quantum-inspired controlled-SWAP (c-SWAP) gate, also called the Fredkin gate~\cite{Fredkin1982}, which is a fundamental building block to realize an SMP protocol. The c-SWAP gate is a three-qubit gate, where the state of one of the qubits determines the swapping of the other two qubits. This gate is also important for the development of quantum computers~\cite{Milburn1989}. Current realizations of the c-SWAP gate include optical~\cite{Garai2014,Mandal2015,Cohen2016}, quantum-optical~\cite{Patel2016,Ono2017},  atomic~\cite{Chauhan2018}, biological~\cite{Orbach2012}, and solid-state systems~\cite{Liu2018}.

The c-SWAP gate in Urrego's implementation uses structured light in the form of optical vortices (e.g., Laguerre-Gauss or Bessel beams ~\cite{Siegman1986,Durnin1987}). The helicity of the optical vortices (right-handed and left-handed) distinguishes between Alice and Bob. The method encodes information in the complex amplitude carried by the spatial modes (akin to quadrature amplitude modulation), allowing phase- and amplitude-encoding. It employs light's polarization to swap the helicity of the vortices (thus, mimicking the swapping of information). 

The realization of optical communication systems using optical vortices has been demonstrated by several proof-of-concept experiments working in free-space and optical fibers~\cite{Gibson:04, Wang2012,Bozinovic2013}. Similar to other degrees of freedom, such as wavelength, complex amplitude, and polarization, we can exploit optical vortex multiplexing to enhance the bit transfer rate~\cite{Wang2016}. The first experiment to demonstrate this concept was reported by Wang et al.~\cite{Wang2012}. They developed a free-space communication system capable of transferring information in the order of Gbit/s.

In this work, we go one step further to the actual implementation of an SMP protocol using optical vortices. We focus our attention on the referee, who has the task of comparing both signals without revealing their content. Our referee consists of a c-SWAP test circuit: two Hadamard gates plus a c-SWAP gate. Figure~\ref{fig:1} shows the quantum circuit implementing the c-SWAP test. The inputs (Alice and Bob) are the states $|\psi \rangle$ and $|\varphi \rangle$. The control qubit enters in the $|0\rangle$ state. The Hadamard gates (H) transform $|0\rangle$ into $(|0\rangle + |1\rangle) /\sqrt{2}$, and $|1\rangle$ into $(|0\rangle - |1\rangle) /\sqrt{2}$. The SWAP is the operation $|\psi\rangle|\varphi\rangle\rightarrow |\varphi\rangle|\psi\rangle$, which is activated if the control qubit is in the $|1\rangle$ state. At the end of the test, the control qubit is measured. After tracing through the execution of the operators, the probability of detecting the control qubit in the state $|0\rangle$ results $(1+|\langle \psi | \varphi \rangle|^2)/2 $. Hence, the outcome $|0\rangle$ has probability $1$ if the states are equal. If the states are different, both outcomes $|0\rangle$ and $|1\rangle$ are possible. Therefore, if we measure $|1\rangle$, we know for sure that the states are different, but if we measure $|0\rangle$, the decision is uncertain. Alternatives methods to realize the referee's task can be found in the references~\cite{Garcia-escartin2013,Lovitz2018,Cincio2018}.

Urrego's experiment incorporates a referee that requires bit-by-bit comparisons and postprocessing the measurements. In addition to the long acquisition times, the experiment exposes the content of each message. Here, we experimentally demonstrate a real-time SMP protocol by incorporating encoding speeds that overcome the detector's integration times, which allows us to compare both messages in a single detection cycle. For such a task, we use a digital micro-mirror device, which allows high-speed spatial light modulations~\cite{Mirhosseini2013}. We also realize a trigger-free detection scheme, which helps to isolate the detector from the laser source. This scheme devises an error-correcting method that reduces the loss of information when the messages arrive between two detection cycles.

Section~\ref{Sec:SWAPtest} describes the theory of our classical c-SWAP test using optical vortices and time-bin multiplexing. Section~\ref{Sec:experiment} details the experimental implementation. Section~\ref{sec:signal_design} explains the analysis and the design of the messages to realize a trigger-free detection scheme and an effective normalization procedure. Section~\ref{Sec:results} shows the experimental results. Finally, section \ref{Sec:conclusions} discusses possible improvements and variations to our method. 

\section{The c-SWAP test using optical vortices}\label{Sec:SWAPtest}

An optical vortex has two distinct characteristics: i) it carries orbital angular momentum (OAM), and ii) it has a helical wavefront comprising a phase singularity and a null intensity on the optical axis~\cite{Allen1992}. Our idea consists of using OAM beams as information carriers, where their helicity (positive or negative) differentiate the two senders. Alice's message is carried by positive OAM modes, whereas Bob's message is carried by negative OAM modes. The information is codified in the complex amplitude accompanying each mode. The beam polarization constitutes the control qubit, where two orthogonal states of polarization represent the $|0\rangle$ and $|1\rangle$ states of the quantum version.   

\begin{figure}[htbp]
\centering
\includegraphics[width=10 cm]{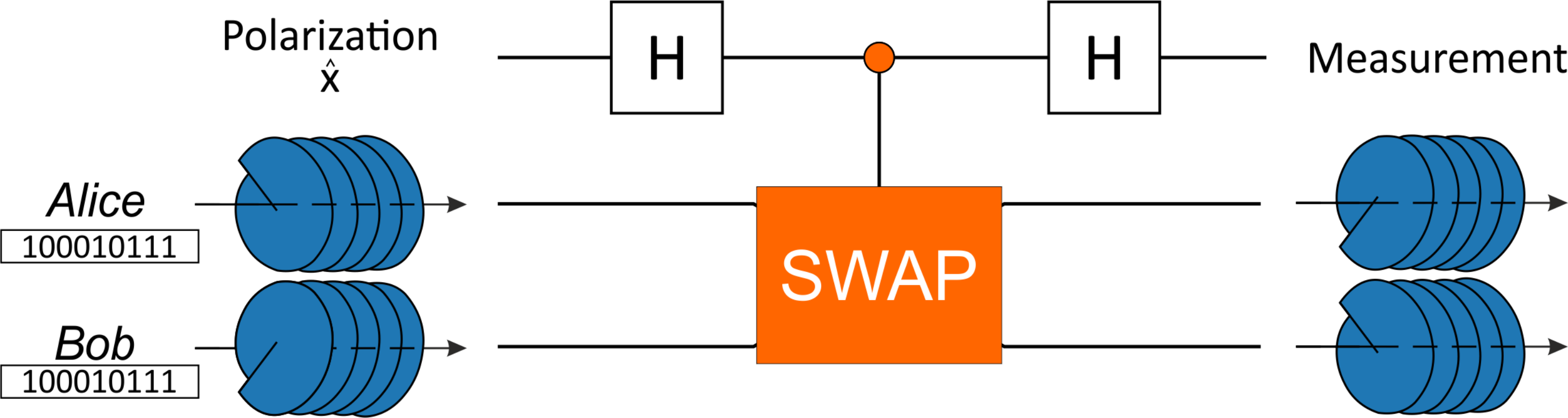}
\caption{Classical c-SWAP test circuit using optical vortices. H stands for Hadamard gate. SWAP is the c-SWAP gate. Alice's message is encoded in the complex amplitude of positive OAM modes, whereas Bob's message is encoded in OAM modes with negative helicity.  
 \label{fig:2}}
\end{figure}

The optical vortices of our choice are Laguerre-Gauss (LG) modes. In a cylindrical coordinate system with $\mathbf{r}=(\mathbf{r}_{\perp},z)=(r,\phi,z)$, an LG beam has an azimuthal phase term $\exp(i m \phi)$, and contains OAM equal to $m\hbar$ per photon (where the mode index $m=0,\pm1,\pm2,\ldots$ is the topological charge, and $\hbar$ is the reduced Planck's constant). In the waist plane $z=0$, the complex amplitude of a single-ringed LG beam (i.e., with radial index equals $0$) is written as
\begin{equation}
    \mathrm{LG}_m (\mathbf{r}_{\perp}) = C_m \left(\frac{r}{w_0}\right)^{|m|} \exp\left( -\frac{r^2}{w_0^2} \right)\exp(i m \phi).
\end{equation}
The constant amplitude $C_m$  normalizes the beam to unit power. The scaling factor $w_0$ is the beam waist. In our protocol, the two senders are LG spatial modes with positive and negative indexes. Modes with positive index correspond to sender 1 (Alice) and modes with negative $m$ correspond to sender 2 (Bob). Owing to their natural orthogonality, LG beams can be used to increase the transmission capacity through multiplexing independent data channels. Here, a channel refers to an electromagnetic wave whose properties remain independent from other channels during propagation~\cite{Torres2012}. We allow the senders to communicate by using multiple channels. 

Figure~\ref{fig:2} shows the classical circuit to realize a c-SWAP test using optical vortices~\cite{Urrego2020}. The information of sender 1 is encoded into $N$ channels with complex amplitudes $A_m$, and the information of sender 2 is similarly encoded into $N$ complex amplitudes $B_m$. The amplitude of the electric field entering the circuit is
\begin{eqnarray}
    \mathbf{E}(\mathbf{r}_{\perp})&=& \sum_{m=1}^{N} \left[A_{m}\mathrm{LG}_{m}(\mathbf{r}_{\perp})+B_{m} \mathrm{LG}_{-m}(\mathbf{r}_{\perp})\right]\mathbf{\hat{x}},\nonumber \\
    &=& \sum_{m=1}^{N} \left[A_{m}u_m+B_{m} v_m\right]\mathbf{\hat{x}}. \label{Eq:input}
\end{eqnarray}
Notice that the protocol does not contain the zeroth-order channel ($m\geq1$). $\mathbf{\hat{x}}$ corresponds to the horizontal polarization state (whose orthogonal polarization is $\mathbf{\hat{y}}$). The Hadamard operation, which consists of a half-wave plate, transforms the input polarization state $\mathbf{\hat{x}}$ to a diagonal state with polarization $(\mathbf{\hat{x}}+\mathbf{\hat{y}})/\sqrt{2}$, and transforms the orthogonal polarization state $\mathbf{\hat{y}}$ into $(\mathbf{\hat{x}}-\mathbf{\hat{y}})/\sqrt{2}$. To simplify the notation, $u_{m}$ and $v_{m}$ describe LG beams with positive and negative mode indexes, respectively [see the second line of Eq.~(\ref{Eq:input})].

It is well-known that a single reflection by a mirror reverses the sense of an optical vortex, i.e., $m \rightarrow -m$~\cite{Gonzalez2006}. We control the SWAP operation by reflecting the beam by an even or odd number of times if the polarization state is $\mathbf{\hat{x}}$ or $\mathbf{\hat{y}}$, respectively. For example, if the light beam entering the c-SWAP has vertical polarization, and contains two senders each carrying one channel, i.e., $\mathbf{E}=\left(A_{1}u_1+B_{1} v_1\right)\mathbf{\hat{y}}$, the c-SWAP operation produces 
\begin{equation}
   A_{1}u_{1}+B_{1}v_{1} \longrightarrow B_{1}u_{1} + A_{1}v_{1},  
\end{equation}
which is equivalent to swapping the information between senders (notice that $u_{-1}=v_{1}$ and $v_{-1}=u_{1}$). On the other hand, if the input beam is $\mathbf{\hat{x}}$-polarized, the c-SWAP realizes the identity transformation. 

The transformations of the input field in Eq.~(\ref{Eq:input}) through the SWAP test circuit (shown in Fig.~\ref{fig:2}) are the following:
\begin{eqnarray}
    \xrightarrow{\text{\,\;\;\quad Input\quad\;}} &\sum_{m=1}^{N} \left[A_{m}u_m+B_{m}v_m\right]\mathbf{\hat{x}},& \label{Eq:inputtocSWAP}\\
    \xrightarrow{\text{Hadamard\,1}} &\sum_{m=1}^{N} \left[ A_{m}u_m+B_{m}v_m \right]\frac{\mathbf{\hat{x}}+\mathbf{\hat{y}}}{\sqrt{2}},&\\
    \xrightarrow{\text{ \,\, c-SWAP\quad}} &\sum_{m=1}^{N} \left[A_{m}u_m+B_{m}v_m\right]\frac{\mathbf{\hat{x}}}{\sqrt{2}}& \nonumber \\ 
    &+ \sum_{m=1}^{N} \left[B_{m}u_m+A_{m}v_m\right]\frac{\mathbf{\hat{y}}}{\sqrt{2}},&\\
    \xrightarrow{\text{Hadamard\,2}} &\sum_{m=1}^{N} (A_m + B_m)(u_m + v_m)\frac{\mathbf{\hat{x}}}{2}& \nonumber \\ 
    &+ \sum_{m=1}^{N} (A_m - B_m)(u_m - v_m)\frac{\mathbf{\hat{y}}}{2}.&
\end{eqnarray}

At the end of the SWAP test circuit, we measure the output powers of each polarization component, which results in
\begin{eqnarray}
    P_x &\propto& \sum_{m=1}^{N} |A_m + B_m |^{2},\nonumber \\
    P_y &\propto& \sum_{m=1}^{N} |A_m - B_m |^{2}.\label{Eq:outpowercomp}
\end{eqnarray}
By observing the above equations, we can conclude that if Alice and Bob send the same message (i.e., $\forall \, m, A_{m}=B_{m}$), we would measure, ideally, $P_y=0$ and $P_x=P_{\mathrm{in}}$, since $P_x+P_y=P_{\mathrm{in}}$. However, in practice, $P_y=C_0$ and $P_x=\eta P_{\mathrm{in}} + C_0 $, where $\eta$ takes into account the detectors' efficiency and losses of the setup, and $C_0$ is the background noise when no input is considered. For the case when both messages are completely different (i.e., $\forall \, m, A_{m} \neq B_{m}$), the result for $P_x$ and $P_y$ depends on the possible values of $A_m$ and $B_m$.

We choose phase encoding to realize a binary system using phases $0$ and $\pi$ and unitary amplitudes. Hence $A_m$,$B_m$ have two possible values $\pm 1$ (a multilevel digital system can be implemented by assigning more phase values). In this agreement, if both messages are different then $A_{m}=-B_{m}$, and we would measure $P_x=0$ and $P_y=P_{\mathrm{in}}$. We notice that Eq.~(\ref{Eq:outpowercomp}) is reminiscent of the Bhattacharyya coefficient~\cite{Fuchs1999}, which is a measure of how distant are two probability distributions. These results corroborate the main objective of the SWAP test: to measure the similarity between Alice's and Bob's messages without revealing their content directly.

For further understanding of the last statement, let us assume that when comparing the two messages we find $n$ identical bit pairs. We define the \textit{overlap} as the percentage of similarity between the two messages as 
\begin{equation}
    \mathrm{Overlap} = (n/N)\times 100\,\%,
\end{equation}
where $N$ is the total number of bits on each message. We remark that the similarity fraction $n/N$ is not a measure of the "distance" between quantum states as stipulated in the quantum version of the SWAP test~\cite{Cincio2018}. However, in a classical system, we consider this number a practical indicator of similarity between digital signals. In order to compute a quantity related to the similarity fraction $n/N$, we define the factor $\gamma$ following reference \cite{Urrego2020} as
\begin{equation}
    \gamma = \frac{P_y - P_x}{P_y+P_x} = - \frac{2\sum_{m=1}^{N} \mathrm{Re}\left[A_{m} B_{m}^{\ast} \right]}{\sum_{m=1}^{N}\left(|A_m|^2+|B_m|^2\right)}.\label{Eq:gammageneral}
\end{equation}
By considering our binary system with values of $\pm 1$, this $\gamma$ factor reduces to
\begin{equation}
    \gamma = 2(n/N) - 1.\label{Eq:gamma_nN}
\end{equation}
Therefore, $\gamma$ is linearly related to the similarity fraction $n/N$. So $\gamma$ is equal to $-1$ for $0\%$ of similarity ($n=0$), and $\gamma=1$ for $100\%$ of similarity ($n=N$). Furthermore, by considering that $P_x + P_y$ is constant and equal to the input power $P_{\mathrm{in}}$, we can express $\gamma$ in Eq.~(\ref{Eq:gammageneral}) as
\begin{equation}
    \gamma = 2(P_{y}/P_{\mathrm{in}}) - 1.\label{Eq:gamma_powerin}
\end{equation}
By comparing Eqs.~(\ref{Eq:gamma_nN}) and (\ref{Eq:gamma_powerin}), it is clear that $P_y=(n/N)P_{\mathrm{in}}$. Thus, we only require to measure $P_y$ to determine the similarity between the signals.  It is true that by measuring both output powers $P_x$ and $P_y$ we can correct power fluctuations of the laser source. Here, however, we apply an active calibration procedure that takes into account errors coming from the laser instabilities, detector efficiency and misalignment.

Additional to OAM multiplexing, we can also consider time-division multiplexing, which consists of dividing the integration time of the detection system into different segments called time bins. Each time bin carries a field of the form given by Eq.~(\ref{Eq:input}). Assuming $N$ OAM channels and $B$ time bins, each message contains $N \times B$ digits per integration time. Let us consider that Alice and Bob messages consist of $N\times B$ digits per integration cycle. The output powers after the SWAP test circuit are now
\begin{eqnarray}
\overline{P}_x &\propto& \tau\sum_{k=1}^{B}\sum_{m=1}^{N} |A_{m,k} + B_{m,k}|^2,\\
\overline{P}_y &\propto& \tau\sum_{k=1}^{B}\sum_{m=1}^{N} |A_{m,k} - B_{m,k}|^2.
\end{eqnarray}
The labels $\{m,k\}$ indicates the $m$-th OAM channel and the $k$-th time bin. $\tau$ is a factor that takes into account the duration of the time bin $\Delta T$.

The overall degree of overlap $\overline{\gamma}$ is
\begin{equation}
    \overline{\gamma} = \frac{\overline{P}_y-\overline{P}_x}{\overline{P}_y+\overline{P}_x} = \frac{\sum_{k=1}^{B} (P_y - P_x)}{B P_{\mathrm{in}}} = \frac{\sum_{k=1}^{B} \gamma_{k}}{B}.
\end{equation}
In a binary system with possible values of $\pm 1$, $\overline{\gamma}$ can be written as
\begin{equation}
    \overline{\gamma} = \frac{\sum_{k=1}^{B} [2(n_k / N)-1]}{B} = \frac{ 2(\overline{n}/N)-B}{B} = 2\left(\frac{\overline{n}}{NB}\right) - 1.
\end{equation}
$\overline{n}$ is the total number of bit pairs that are equal considering each message has a total number of bits $N \times B$. Therefore, $\overline{n}/NB$ is the new similarity fraction between the messages. As previously, since  $\overline{P}_{x} + \overline{P}_{y} = \overline{P}_{\mathrm{in}}$ is constant, we can write 
\begin{equation}
    \overline{\gamma} = 2\left(\frac{\overline{P}_y}{\overline{P}_{\mathrm{in}}}\right)-1.
\end{equation}

We notice that now, we can still measure the similarity fraction using a single detector for the $y$ polarization component. However, we need to measure the integrated power $\overline{P}_{x}+\overline{P}_y = \overline{P}_{\mathrm{in}}$ (the power integrated over all time bins). In order to find this integrated power without using the second detector, we consider a reference signal. This reference signal corresponds to the case where $A_{m,k}=-B_{m,k}\; \forall \{m,k\}$, therefore $\overline{P}_{x}=0$ and $\overline{P}_{y}=\overline{P}_{\mathrm{in}}$. We also use a second reference signal where $A_{m,k} = B_{m,k}\; \forall \{m,k\}$ and therefore $\overline{P}_x=\overline{P}_\mathrm{in}$ and $\overline{P}_y$ ideally equals $0$. In practice, the minimum and maximum values are not achieved (due to background illumination, detectors' efficiency and imperfections in the experiment), so we use these the measurements from the reference signals to calibrate the system.

\section{Experimental implementation}\label{Sec:experiment}

For this implementation to work, we require a fast light modulator to overcome the detector's integration time. Liquid crystal (LC)-based spatial light modulators (SLMs) are typically used to shape laser beams into ones carrying optical vortice. However, their low frame rate ($\sim$Hz) represents a limitation for high-speed applications. Digital micro-mirror devices (DMDs) have emerged as a solution to this problem~\cite{Turtaev2017}, offering frame rates of kHz. Here, the DMD is the key element for implementing the SMP protocol. We use it for both beam shaping and time-division multiplexing.

\begin{figure*}[htbp]
\centering
\includegraphics[width=\textwidth]{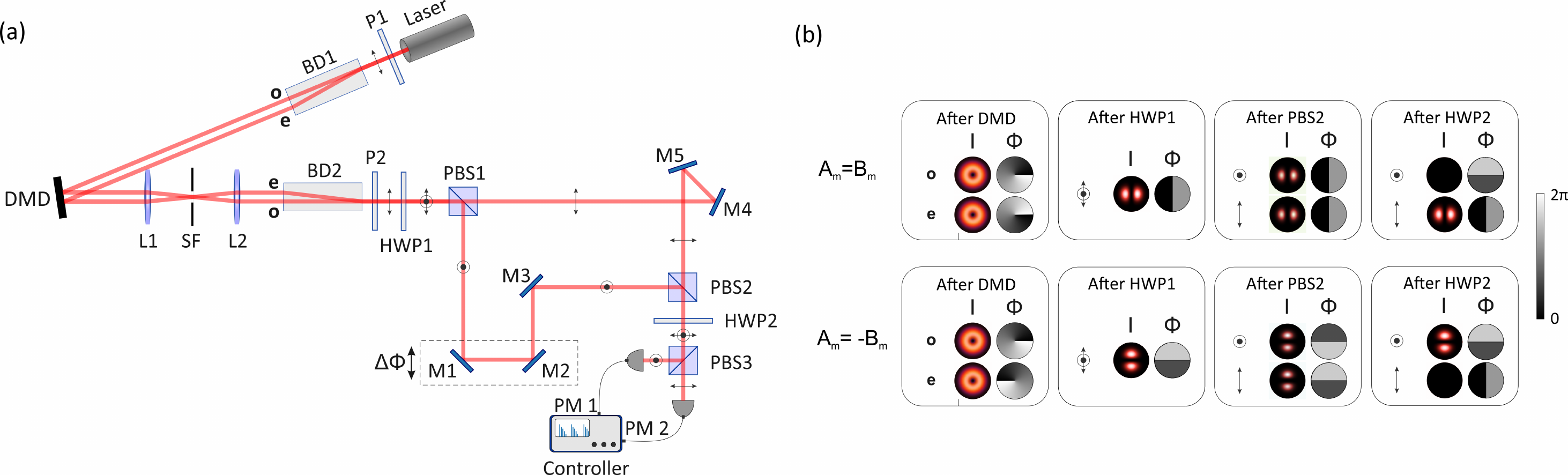}
\caption{Experimental setup. (a) P1 and P2, linear polarizers; BD1 and BD2, beam displacers; DMD, digital micro-mirror device; L1 and L2; lenses; SF, spatial filter; HWP1 and HWP2, half-wave plates; PBS1, PBS2 and PBS3; polarizing cube beamsplitters; M1-M5, mirrors; PM1 and PM2, power meters. (b) Beam intensity, phase distribution and polarization components at different stages of the experiment. 
 \label{fig:3}}
\end{figure*}

Our SWAP test implementation uses a Mach-Zehnder (MZ) interferometer. It has been shown that the angular sensitivity in the interference of vortex beams scales linearly with the topological charge~\cite{Jha2011}. Even with our calibration procedure, we found our interferometer to be highly sensitive when considering up to three OAM channels per sender. In fact, this issue also limits Urrego's implementation in reference \cite{Urrego2020}, where they only consider a maximum of two OAM channels. Consequently, we only consider time-division multiplexing using one OAM channel per sender. In addition, to adopt the formulation presented in section \ref{Sec:SWAPtest}, we consider a binary system where the complex amplitudes $A_{1},B_{1}$ can have values of $\pm 1$. 

Let us also note that this experimental demonstration requires two types of detectors: a slow and a fast detector. The slow detector defines the detection cycle of our SMP protocol, while the fast detector is used to measure the switching rate of our DMD. For both types of detectors (slow and fast), we use the same photodiode power sensor (Thorlabs S120C). This sensor has a time resolution of about $1$~$\mu$s. Our fast detector consists of the analog output from this sensor, which is visualized using an oscilloscope. The slow detector is the digital output, which is visualized using the power meter interface. The digital output time resolution depends on several factors: the analog-to-digital conversion rate, the USB transfer rate, and the computer integration time. After several instances of the experiment (more details in the following sections), we estimate that the time resolution of the slow detector varies between $4$ and $8$~ms, with an average of $5.32$~ms. 

Considering the above discussion, we set the integration time of our protocol to be $5.32$~ms. This integration time is called the average-time-gap of the receiver and is labeled as $t_g^r$. Since our DMD has a frame rate of $f_r = 9.5$~kHz, in principle, we can achieve $t_g^r \times f_r \approx 50$ time bins per message, each time bin lasting about $105$~$\mu$s. However, while its apparent that the factor limiting  the number of time bins is the DMD's frame rate, there is an additional technical detail to consider: the DMD's memory capacity. To avoid saturating the DMD's memory, but without loosing our objective, we consider $40$ digits per sender using a modulation rate of $7.5$ kHz.

\subsection{Setup description}
Figure~\ref{fig:3} shows a schematic of our experiment. We use a linearly-polarized  He-Ne laser beam (Thorlabs HNL020LB) with wavelength $\lambda=632.8$~nm and power $P=2$~mW. The beam is split directly after the laser by means of a calcite beam displacer (Thorlabs BD40). The beam displacer is oriented at $45^{\circ}$ regarding the $\mathbf{\hat{x}}$-polarization direction, which is chosen parallel to the optical table. Additionally, we use a linear polarizer (P1) to fine-tune the splitting to be as close to $50/50$ as possible. The beams exiting the beam displacer have orthogonal polarizations, one is the ordinary polarized ($\mathbf{o}$), and the other is the extraordinary polarized ($\mathbf{e}$).  Those two spatially separated beams can now be shaped independently by the DMD with two different holograms. The polarization-independent nature of the DMD enables no further adjustment of the beam's polarization. 

The DMD (DLP LightCrafter 6500 manufactured by Texas Instruments) is an amplitude-only spatial light modulator with a screen resolution of $1920 \times 1080$ pixels [px], and $8$~$\mu$m pixel pitch capable of manipulating the beam up to $9.5$~kHz. It is mounted on a $\mu$m stage in order to displace it at high precision. Its mount was 3D printed in house. The holograms displayed on the DMD correspond to two LG modes with opposite topological charges. The two beams are shaped spatially separated, simulating the two senders Alice and Bob having one OAM channel each. The modulated beams pass through a 4f system (lens L1 and L2) with an aperture (SF) to isolate, and Fourier transform the first diffraction order. A second beam displacer (BD2) recombines the two beams. The linear polarizer (P2) passes the $\mathbf{\hat{x}}$-polarized component. 

Up to this point, the beam carries the messages of Alice and Bob, which are emulated with the DMD, according to Eq.~(\ref{Eq:inputtocSWAP}). Alice encodes her message with left-handed vortices (positive mode indexes $m>0$), and Bob with right-handed vortices (negative mode indexes $m<0$). Our control qubit, i.e., light's polarization, is in the $|0\rangle$-state, corresponding to horizontal polarization.

The Hadamard gates consist of half-wave plates (HWP1 and HWP2) with the fast axis oriented $22.5^{\circ}$ with respect to the horizontal polarization. The c-SWAP gate consists of an unbalanced and polarization-sensitive MZ interferometer. The first polarizing beam-splitter (PBS1) separates the beam into horizontal ($\mathbf{\hat{x}}$-polarization) and vertical ($\mathbf{\hat{y}}$-polarization) polarization states. The $\mathbf{\hat{x}}$-polarized [$\mathbf{\hat{y}}$-polarized] component propagates through the upper [lower] arm and experiences an even [odd] number of reflections, which preserves [reverses] the helicity of the optical vortices. The moving trombone system of mirrors (M1 and M2) in the lower arm compensates the phase difference between the beams reaching the second PBS (PBS2). The beam exiting HWP2 goes to the detection stage, which consists of a PBS (PBS3) and two power meters (PM1 and PM2).

Figure~\ref{fig:3}(b) shows the beam intensity and phase distribution at the four main stages of the experiment for the case $A_1 = B_1$ in the upper row and for $A_1 = -B_1$ in the lower row, respectively. The first column shows the beam after the DMD in its two spatially separated polarization states $\mathbf{o}$ and $\mathbf{e}$: pure LG modes with $m = \pm 1$. The second column sketches the beam as it enters the MZ interferometer. The third column shows the beam after the c-SWAP. Note the resulting reflection in the phase distribution for the vertical polarization, which propagates through the lower interferometer arm and is subject to the OAM swap. The fourth column shows the beams intensity and phase distribution after the second Hadamard gate (HWP2).   

\subsection{Frequency determination}
\begin{figure}[htbp]
\centering
\includegraphics[width=\columnwidth]{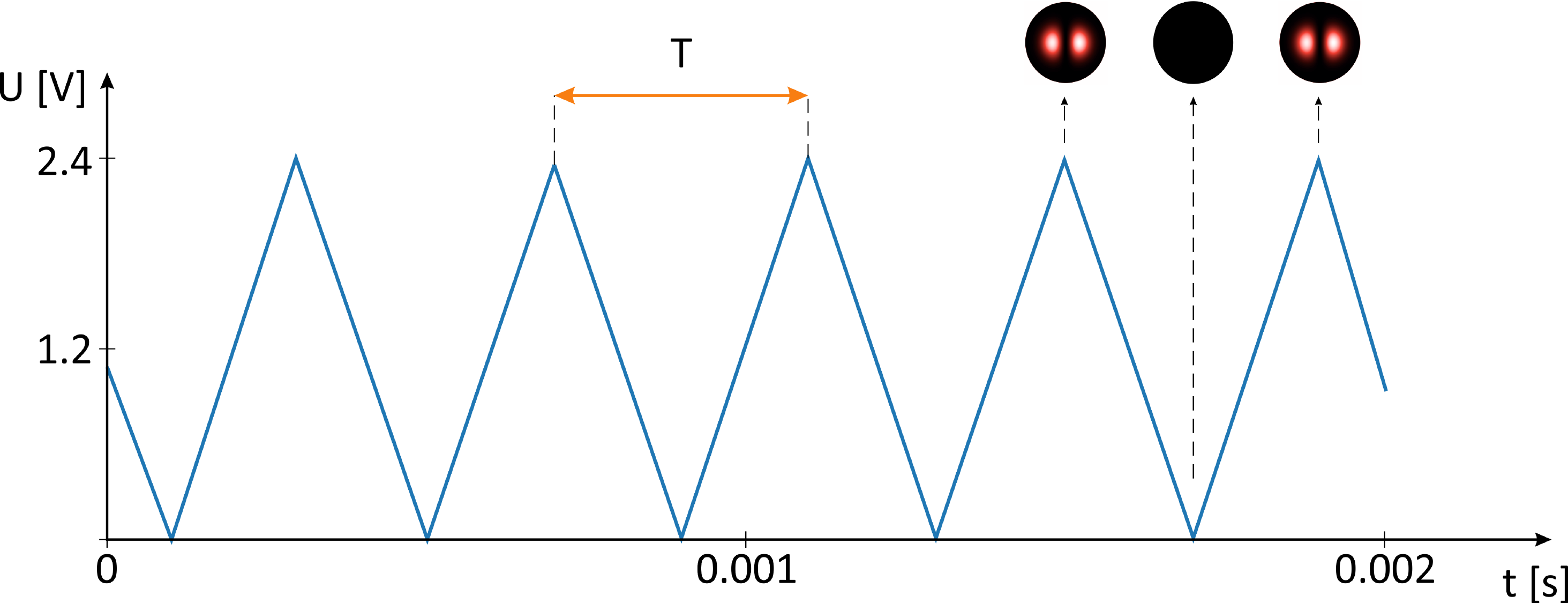}
\caption{This schematic sketches the method we used to determine the modulation rate of our DMD. Switching between the LG-$\pm m$-hologram and the pause-hologram one gets a signal in the shown form. We can calculate the time-difference $T$ between two same holograms, which enables to calculate the effective exposing time of one hologram, and from that the frequency of the DMD modulation.
 \label{fig:freqdet}}
\end{figure}

One of the power meters from the experimental setup is connected to an oscilloscope to determine the DMD's frequency modulation. A similar method appears in reference \cite{Mirhosseini2013}. We display two holograms on the DMD in a continuous mode, switching between those two without an offset. First of the displayed holograms was the one generating a superposition of LG modes, the second hologram corresponds to the pause hologram, which redirects the beam outside the setup, resulting in a dark signal (no light detected). Figure~\ref{fig:freqdet} shows a schematic of the method. According to the schematic, the DMD frame rate is $2/T$, where $T$ is the time separation between two maxima (notice that the zero detected power is equal to the pause hologram). An actual photo of the oscilloscope measurement appears in the next subsection (see Fig.~\ref{fig:pre_adjust}).

\subsection{Preliminary adjustments}
In order to check that the setup is aligned and ready to perform the SMP protocol, we have to ensure that we get opposite power maxima in the $\mathbf{\hat{x}}$ and $\mathbf{\hat{y}}$ outputs, for the cases where $A_1 = B_1$ and $A_1 = -B_1$. We expect a maximum in $\mathbf{\hat{x}}$- polarization and a minimum in $\mathbf{\hat{y}}$-polarization for the case $A_1 = B_1$, while for $A_1 = -B_1$ the measurement must result in a minimum in $\mathbf{\hat{x}}$-polarization and a maximum in $\mathbf{\hat{y}}$-polarization. To check if we are ready to measure, we alternately display the two holograms, i.e., the hologram generating the LG-beam superposition with $A_1 = B_1$ and the hologram for $A_1 = -B_1$. Each hologram is exposed for $\Delta T = 1000$~$\mu$s. Doing so, we expect alternating maxima and minima in both power meters (PM1, PM2). If PM1 measures the power in $\mathbf{\hat{x}}$-polarization and PM2 the power in $\mathbf{\hat{y}}$-polarization, we expect a maximum in PM1 and a minimum in PM2 for the holograms with $A_1=B_1$ and vice versa for $A_1 =-B_1$.

\begin{figure}[htbp]
\centering
\includegraphics[width=\columnwidth]{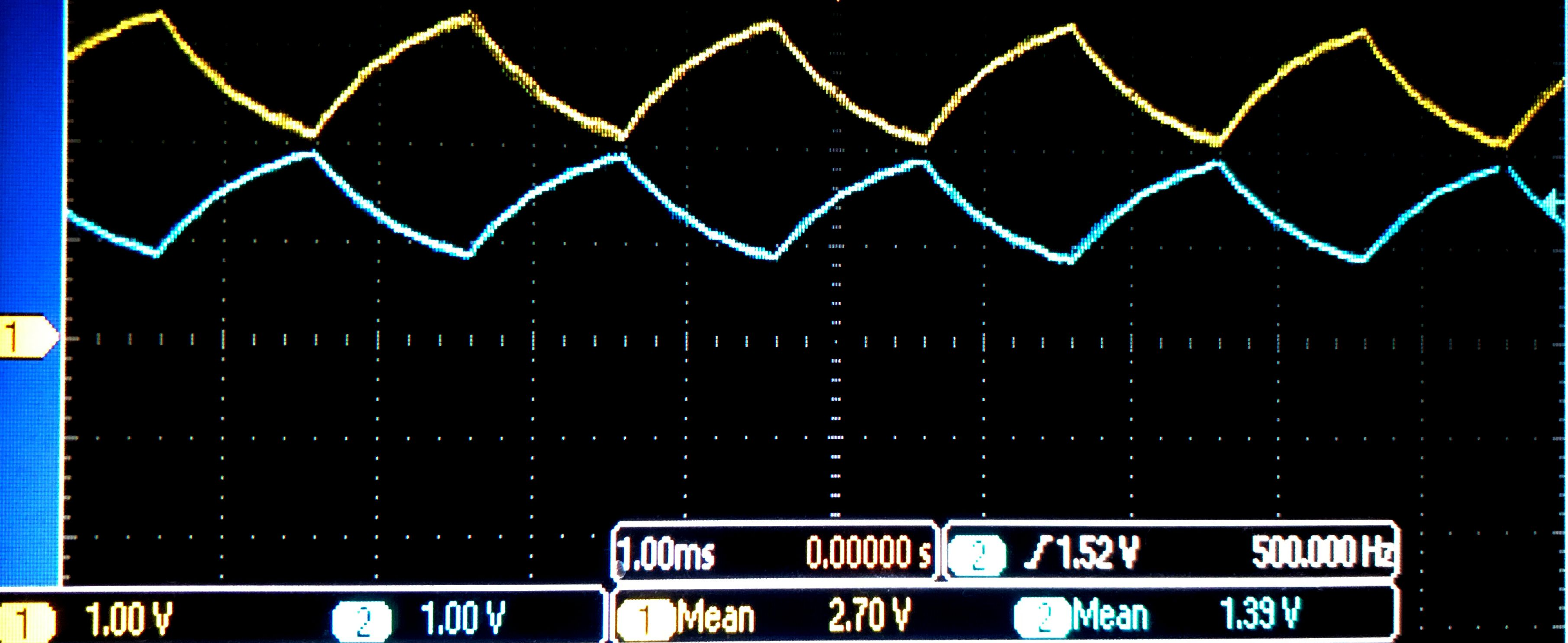}
\caption{Preliminary adjustments. PM1 and PM2 voltage for alternating holograms ($A_1 = B_1$ and $A_1 = - B_1$). In this example, the exposure time of each hologram is $1000$~$\mu$s. Notice that the frequency detected by the oscilloscope ($500$~Hz) is half the frame rate of the DMD ($1$~kHz). 
 \label{fig:pre_adjust}}
\end{figure}

Figure~\ref{fig:pre_adjust} shows a photo of the oscilloscope during this interference test. We can see that when PM1 has a maximum PM2 has a minimum and vice versa, which means that the setup is well-aligned and we can start to measure. Note that the frequency determined by the oscilloscope equals $500$~Hz, which matches the expectation when exposing the holograms for $1000$~$\mu$s. It takes $2000$~$\mu$s to get back from the hologram $A_1 = B_1$ to itself, and therefore the frequency found is $500$~Hz. However, the modulation rate of the DMD is doubled.

\section{Design, analysis and processing} \label{sec:signal_design}
The overall most challenging task of the experiment is the designing of a measurement protocol that fitted our given measurement devices. We work with a continuous laser beam and not a pulsed laser. Thus the measuring and sending of our signal cannot be triggered simultaneously; in other words, it cannot be synchronized. This leads to the issue that the integration time of our receiver (which we call the time-gap-receiver $t_g^r$), i.e., the referee in the form of the power meters (PM1 and PM2), does in general not align with the sending time of Alice and Bob (which we call the time-gap-sender $t_g^s$). Alice and Bob start to send their encoded message at the same time. The message is sent bit by bit over the time interval $t_g^s$, while the referee measures one data point per $t_g^r$. 

\begin{figure}[htbp]
\centering
\includegraphics[width=10 cm]{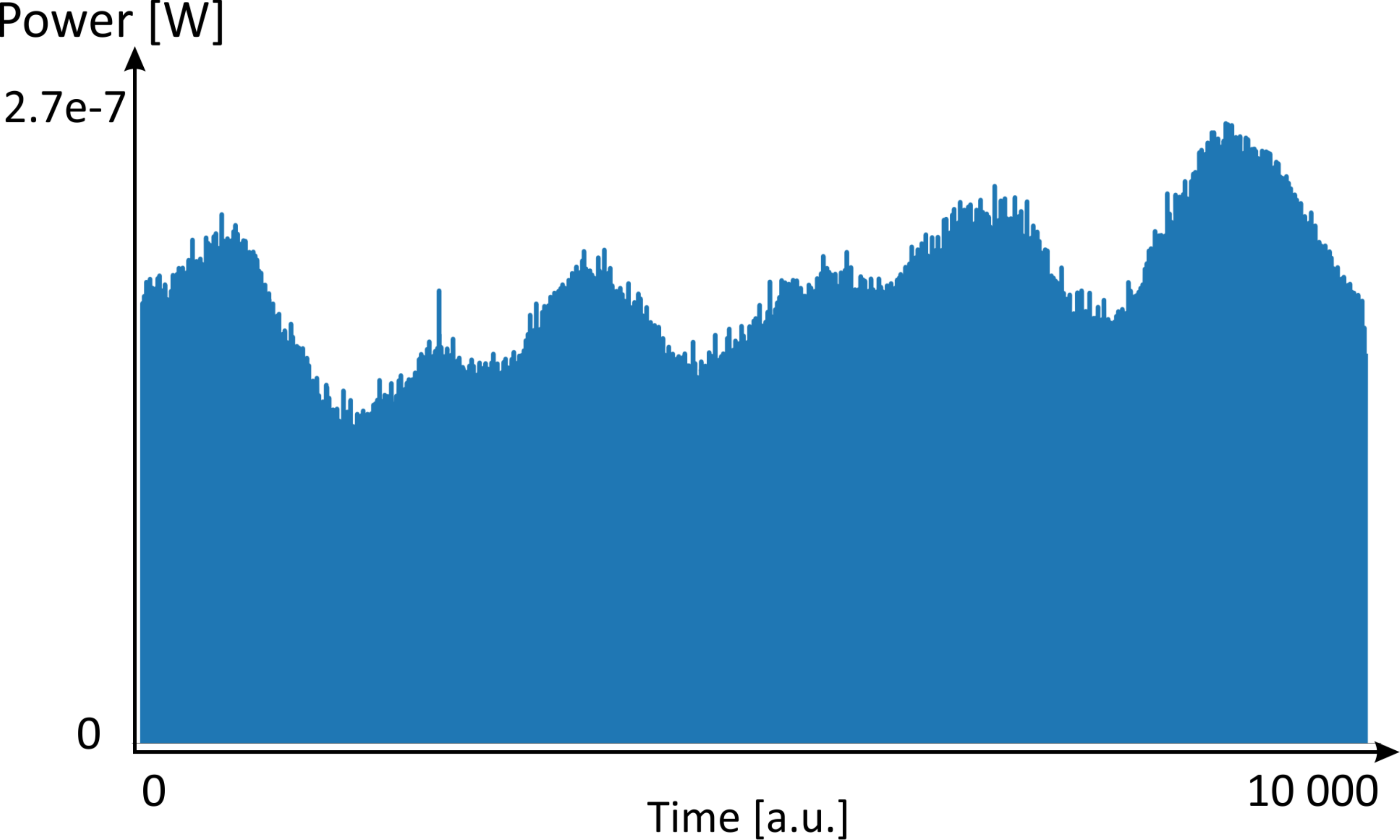}
\caption{Variation of the detected power over one minute time interval. Data taken from the power meter computer interface, which are analyzed and displayed using a custom Python script. 
 \label{fig:signal_var}}
\end{figure}

\subsection{Time-gap conditions}
Two conditions have to be achieved to ensure that one single data point represents one entire message-comparison measurement:
\begin{enumerate}
    \item[(a)]  $t_g^s \overset{!}{=} t_g^r$, the sender and receiver time-gap have to be equal. While the time-gap of the receiver $t_g^r$ is not changeable, we can adapt the time-gap of the sender $t_g^s$ by changing the exposure time of the holograms representing the message-bits. Therefore, the first step is to determine $t_g^r$, which is obtained by analyzing a random data set. We found that $t_g^r$ varies between $4$ and $8$ ms, with an average of $5.32$~ms. Also, the measurement devices have the following information:
    \begin{itemize}
        \item Powermeter S120C from Thorlabs: resolution $<1~\mu$s.
        \item Powermeter console PM100A from Thorlabs: data point is converted to digital value $\approx$ every $0.3$ ms.
        \item USB transfer rate: $\approx$ $3$ ms.
        \item Computer integration time: unknown.
    \end{itemize}
    
    \item[(b)] $t_g^s$ and $t_g^r$ have to start at the same time. If this condition is not satisfied, the detector will cut or ignore part of the message, resulting in an erroneous measurement. Since the DMD and the detector both work in continuous mode, this condition can not be satisfied. Therefore, we had to design a message/measurement protocol to eliminate errors from the lack of sender-receiver synchronization.
\end{enumerate}

\subsection{Calibration}
Additional to the above conditions, the measurements are exposed to errors due to experimental imperfections, mechanical vibrations, phase and intensity fluctuations, background illumination, and detector sensitivity (see Fig.~\ref{fig:signal_var}). We employ two reference message comparisons to incorporate all those factors into the system calibration for each experimental iteration. One comparison corresponding to $100\%$ overlap and another to $0\%$. The former sets the maximum detected power, whereas the latter sets the minimum. Therefore, we calibrate the system using the power measurement of both references.

\begin{figure}[htbp]
\centering
\includegraphics[width=\columnwidth]{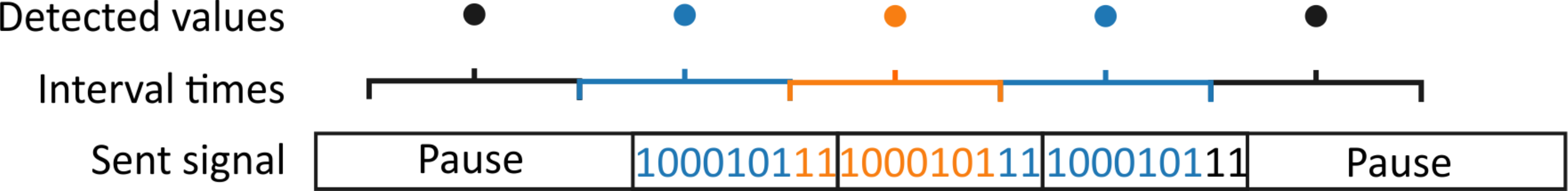}
\caption{ An exemplary message with $N=9$ bits is sent three times in sequence. The time-gaps of the receiver matches the criterion $t_g^s = t_g^r$, however $t_g^s$ and $t_g^r$ are misaligned. Nevertheless, the middle detected value (orange dot) will always account for one complete message comparison, while its two neighbours (blue dots) might contain part of the pause-signal (black dots). 
 \label{fig:signal_const}}
\end{figure} 

\begin{figure}[htbp]
\centering
\includegraphics[width=\columnwidth]{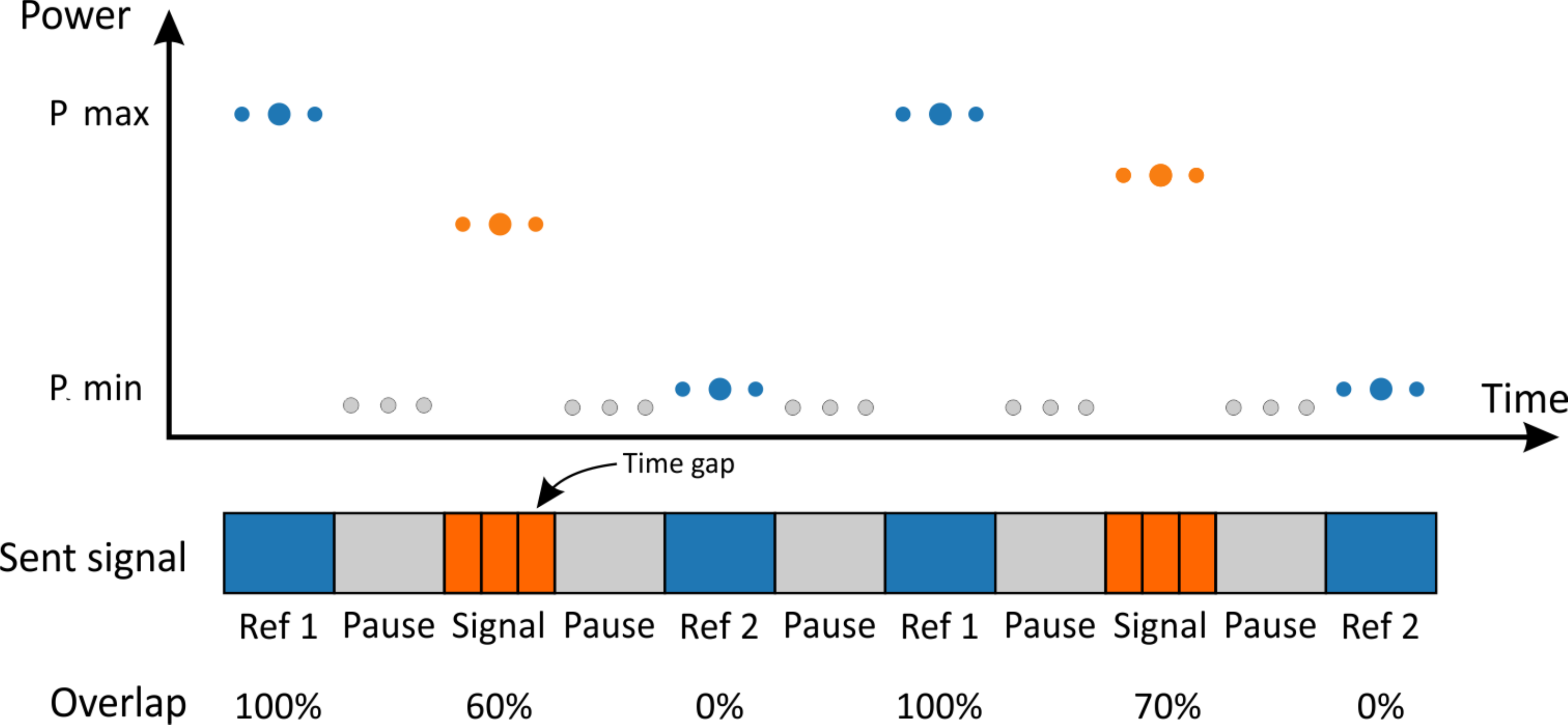}
\caption{Measurement-design protocol with a signal of $60\%$ and $70\%$ overlap. Each signal was sandwiched between reference signals Ref 1 and Ref 2, corresponding to $100\%$ and $0\%$ of overlap, respectively. \label{fig:sig_complet}}
\end{figure}

\subsection{Message and measurement protocol}
Our message/measurement protocol has the following steps:

\begin{enumerate}
    \item Determine the average-time-gap of the receiver. In our case $t_g^r = 5.32$~ms.
    \item Consider $t_g^s = t_g^r$ and determine the number of time bins $N$ according to the frame rate of the DMD. Therefore, $t_g^s=N\cdot \Delta T$, where $\Delta T$ corresponds to the exposing time of the holograms (equivalent to the duration of each time bin).
    \item Alice and Bob send their message $3$ times in sequence without pause. Since the SMP protocol does not require to compare the messages in sequential order, by employing this strategy the second measurement represents one complete message comparison even if $t_g^s$ and $t_g^r$ are misaligned, as illustrated in Fig.~\ref{fig:signal_const}.
    \item Include a $100\%$ overlap hologram $A_1 = B_1$ before and a $0\%$ overlap hologram $A_1 = -B_1$ after each actual message, both separated from the message string by three time gaps, i.e., $3t_g^r$. The reference holograms are also displayed for the time of three time-gaps.
    \item The message-comparison data-point is then locally normalized regarding the two reference peaks Ref 1 and Ref 2. The normalized power $P_n$ is given by
    \begin{equation}
    P_n = \frac{\mathrm{Middle\,Data\,Point} - \mathrm{Min(Ref1,Ref2)}}{|\mathrm{Ref1}-\mathrm{Ref2}|}.
\end{equation}
The $\mathrm{Min}(\cdot)$ function protects the algorithm from migration of power. This may happen if local air fluctuations will modify the optical path difference between the two arms of the interferometer. In that case, the maximum power for the completely overlapped signals detected at one power meter can migrate to the other one. Function $\mathrm{Min}(\cdot)$ keeps the normalization procedure insensitive to that. As long as reference signals represent $0\%$ and $100\%$ of overlap, the analysis algorithm will provide the right normalized power value.
\end{enumerate}

\begin{figure}[htbp]
\centering
\includegraphics[width=\columnwidth]{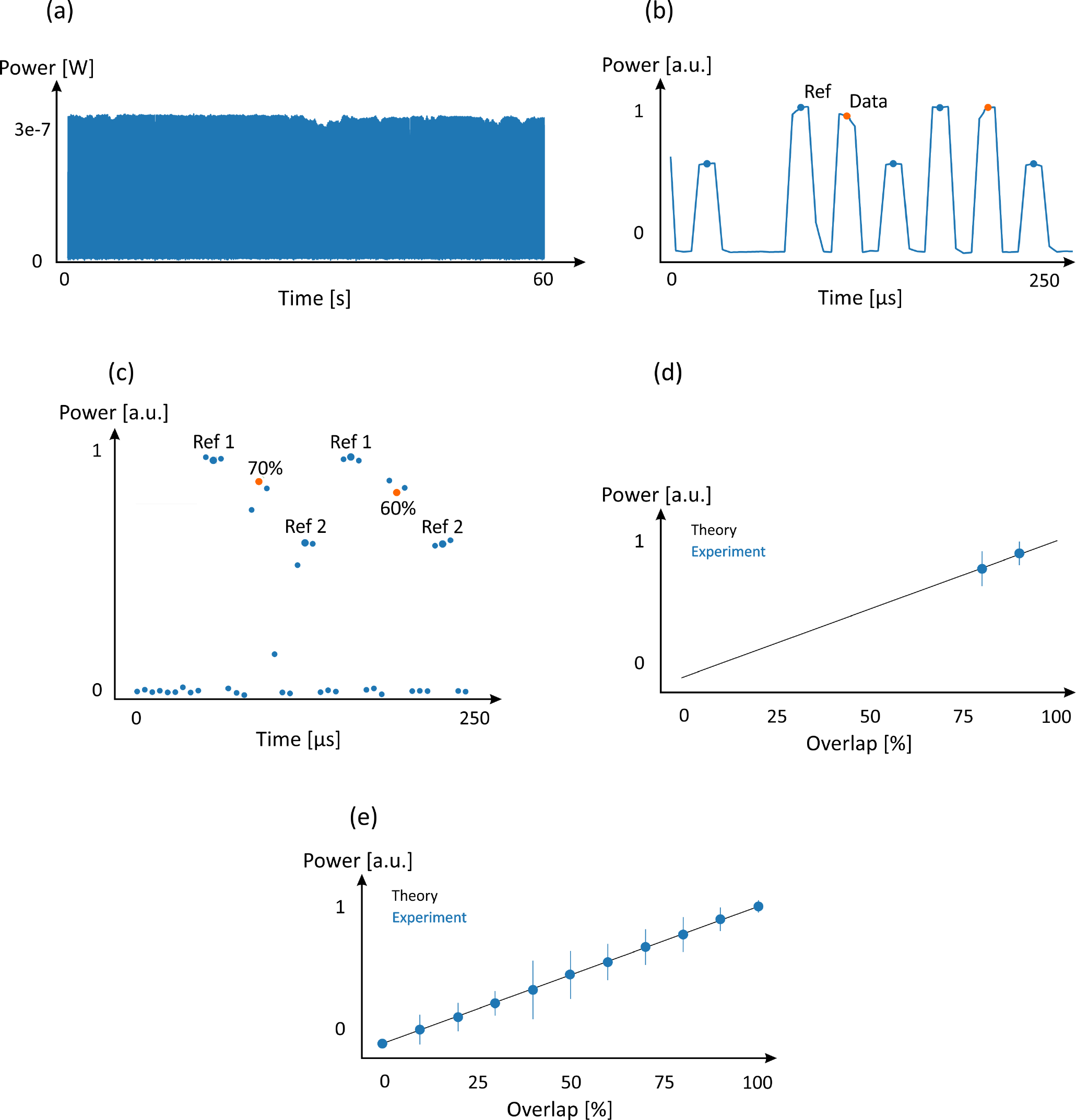}
\caption{Final results of the SMP protocol. (a)-(e) give some insight into the analysis for a $60\%$ and $70\%$ overlap SMP protocol measurement. (f) presents the joined data for all $0\%$-$100\%$ overlap measurements. The data are analyzed and plotted using a custom Python script. See section \ref{Sec:results} for more details.
 \label{fig:results}}
\end{figure}

Figure~\ref{fig:sig_complet} shows a schematic of a final measurement-design including two SMP protocols with different percentages of overlap between the messages sent by Alice and Bob: in the first one, Alice and Bob send two messages which have a $60\%$ overlap and in the second one, they send messages with $70\%$ overlap. The dots represent the data-points.

Once the measurement protocol is designed (consisting of the bit-strings representing Alice's and Bob's messages, the $100\%$ and $0\%$ references and the pauses), it is then saved as txt-file and loaded to the DMD software. For the experiment, the protocol was put into a loop. Thus the last pause of the protocol had to be longer in order to separate the measurement loops. The analysis algorithm finds the loops where the time-gap of the sender equals the time-gap of the receiver for all data-points within this loop, i.e., $t_g^s \approx t_g^r \, \forall$ data-points. It allows for max two misaligned time-gaps per loop and discards all other measurement loops where the alignment between receiver and sender is not given. As $t_g^r$ varies a lot, the shorter the measurement protocol the higher the chance that $t_g^s \approx t_g^r \, \forall$ data-points. This is the reason why we decided to send maximal two different messages-overlap strings with references per measurement-protocol.

To give an idea of how the analysis works, the case shown in Fig.~\ref{fig:sig_complet} can be explained in the following way. This part of the protocol contains two different bit-strings representing the case of $60\%$ and $70\%$ overlap between the messages of Alice and Bob, respectively. Each is sent with reference peaks, and each peak is divided by a $3t_g^r$-long pauses. Thus, one complete protocol will lead to 6 peaks and 5 pauses of which each results in 3 data-points. The algorithm first check, whether there are 6 peaks, then if there are $11 \cdot 3 = 33$ data-points and in order to check that $t_g^s \approx t_g^r \, \forall$ data-points it will check whether $6\cdot3\pm 1=18\pm 1$ data-points are higher than a defined minimum value and $5\times 3 \pm 1 = 15\pm 1$ are lower. Only if all these criteria are met, the measurement loop is considered and the middle dots are taken into account to provide the result. Measurement loops that do not meet the conditions are neglected.   

\section{Results}\label{Sec:results}
Figure~\ref{fig:results} presents the final results of the SMP protocol of two $40$-bit messages sent in time-bins for $0\%$, $10\%$, $20\%$,$\ldots$, $100\%$ overlap, respectively. The receiver time-gap (the referee) was defined to be $t_g^r = 5.32$~ms. Therefore, one hologram was exposed for $5320\mu s /40 = 133 \mu$s, which corresponds to a beam modulation rate of $7.52$~kHz. Each separate measurement contained maximal two different message passing protocols.

In order to gather sufficient data to analyze the statistic of our experiment,  the measurement protocol is put into a loop and measured for one minute. This is done for $0\%$ to $100\%$ overlap of the messages with $10\%$ steps. All data points are normalized regarding their references, and the average is calculated from the normalized values as well as their standard deviations.

Figure \ref{fig:results}(a) plots the raw data for the $60\%$ and $70\%$ overlap protocol measurement put into a loop for one minute. Figures \ref{fig:results}(b) and (c) show one loop which matched the conditions given in section \ref{sec:signal_design}A, thus where the sending time-gap $t_g^s$ overlapped with the receiver time-gap $t_g^r$. In Fig.~\ref{fig:results}(d), the calculated average of the normalized powers for the first $60$ measurement loops, which matched the criteria, is plotted with its standard deviation for $60\%$ and $70\%$ overlap. Also, the theoretical line is plotted and shows that the experimental data corresponds to the theoretical expectations. Finally, Fig.~\ref{fig:results}(e) plots the normalized power average with standard deviation for all $0\%$-$100\%$ overlap measurements. This final result clearly demonstrates that we can successfully determine the similarity of two messages without unveiling the actual messages with up to $10\%$ exactness using optical vortices.

\section{Discussion and conclusions}\label{Sec:conclusions}

It is worth mentioning that our system has an important symmetry regarding the OAM channels: one channel contains only OAM modes with positive $m$ indexes, while the other channel contains negative modes. Due to this symmetry with respect to the sign of $m$, we can also implement the c-SWAP operation with the help of a controlled-NOT (c-NOT) gate. However, note that the c-NOT operation aims at reversing channels, not at exchanging information. If the sign of $m$ represented information, we would be implementing a polarization-controlled NOT operation~\cite{Slussarenko2010}. One could have chosen two channels where $m_1 \neq - m_2$. In this case, the c-SWAP operation would have to be implemented in a different way.

In summary, we proved the overlap of two $40$-bit messages encoded in the complex amplitude of LG modes of order $m\pm 1$ within one single acquisition cycle and one data-point, up to a resolution of $10\%$ overlap, without revealing the content of the messages. We have demonstrated an SMP protocol by implementing a SWAP test using optical vortices. Of course, there is room for improvement. The message capacity can be increased using more OAM channels, using more time-bins (employing a faster DMD), or incorporating multilevel digits (instead of bits). This work takes part in the efforts to incorporate the opportunities offered by optical vortices into a reliable communication system.

\section*{Funding}
Consejo Nacional de Ciencia y Tecnolog\'{i}a (CONACYT) (257517, 280181, 293471, 295239, APN2016-3140); Polish Ministry of Science and Higher Education ("Diamond Grant") (DIA 2016 0079 45); Nacional Science Centre (Poland) (UMO-2018/28/T/ST2/00125).

\bibliographystyle{unsrt}
\bibliography{references}

\begin{thebibliography}{10}

\bibitem{Buhrman2001}
Harry Buhrman, Richard Cleve, John Watrous, and Ronald de~Wolf.
\newblock {Quantum Fingerprinting}.
\newblock {\em Physical Review Letters}, 87:167902, 2001.

\bibitem{Arrazola2014}
Juan~Miguel Arrazola and Norbert L{\"{u}}tkenhaus.
\newblock {Quantum fingerprinting with coherent states and a constant mean
  number of photons}.
\newblock {\em Physical Review A}, 89:062305, 2014.

\bibitem{Xu2015}
Feihu Xu, Juan~Miguel Arrazola, Kejin Wei, Wenyuan Wang, Pablo Palacios-Avila,
  Chen Feng, Shihan Sajeed, Norbert L{\"{u}}tkenhaus, and Hoi-Kwong Lo.
\newblock {Experimental quantum fingerprinting with weak coherent pulses}.
\newblock {\em Nature Communications}, 6:8735, 2015.

\bibitem{Guan2016}
Jian-Yu Guan, Feihu Xu, Hua-Lei Yin, Yuan Li, Wei-Jun Zhang, Si-Jing Chen,
  Xiao-Yan Yang, Li~Li, Li-Xing You, Teng-Yun Chen, Zhen Wang, Qiang Zhang, and
  Jian-Wei Pan.
\newblock {Observation of Quantum Fingerprinting Beating the Classical Limit}.
\newblock {\em Physical Review Letters}, 116:240502, 2016.

\bibitem{Urrego2020}
Daniel~F Urrego, Dorilian Lopez-Mago, Ver{\'{o}}nica
  Vicu{\~{n}}a-Hern{\'{a}}ndez, and Juan~P Torres.
\newblock {Quantum-inspired Fredkin gate based on spatial modes of light}.
\newblock {\em Opt. Express}, 28:12661, 2020.

\bibitem{Fredkin1982}
Edward Fredkin and Tommaso Toffoli.
\newblock {Conservative logic}.
\newblock {\em International Journal of Theoretical Physics}, 21:219--253,
  1982.

\bibitem{Milburn1989}
G.~J. Milburn.
\newblock {Quantum optical Fredkin gate}.
\newblock {\em Phys. Rev. Lett.}, 62:2124--2127, 1989.

\bibitem{Garai2014}
Sisir~Kumar Garai.
\newblock {A novel method of developing all optical frequency encoded Fredkin
  gates}.
\newblock {\em Opt. Commun.}, 313:441--447, 2014.

\bibitem{Mandal2015}
Dhoumendra Mandal, Sumana Mandal, and Sisir~Kumar Garai.
\newblock {Alternative approach of developing all-optical Fredkin and Toffoli
  gates}.
\newblock {\em Opt. Laser Technol.}, 72:33--41, 2015.

\bibitem{Cohen2016}
Eyal Cohen, Shlomi Dolev, and Michael Rosenblit.
\newblock {All-optical design for inherently energy-conserving reversible gates
  and circuits}.
\newblock {\em Nat. Commun.}, 7:11424, 2016.

\bibitem{Patel2016}
Raj~B. Patel, Joseph Ho, Franck Ferreyrol, Timothy~C. Ralph, and Geoff~J.
  Pryde.
\newblock {A quantum Fredkin gate}.
\newblock {\em Sci. Adv.}, 2:e1501531, 2016.

\bibitem{Ono2017}
Takafumi Ono, Ryo Okamoto, Masato Tanida, Holger~F. Hofmann, and Shigeki
  Takeuchi.
\newblock {Implementation of a quantum controlled-SWAP gate with photonic
  circuits}.
\newblock {\em Sci. Rep.}, 7:45353, 2017.

\bibitem{Chauhan2018}
Anil~Kumar Chauhan and Asoka Biswas.
\newblock {Atomic swap gate, driven by position fluctuations, in dispersive
  cavity optomechanics}.
\newblock {\em J. Mod. Opt.}, 66:438--447, 2019.

\bibitem{Orbach2012}
R.~Orbach, F.~Remacle, R.~D. Levine, and I.~Willner.
\newblock {Logic reversibility and thermodynamic irreversibility demonstrated
  by DNAzyme-based Toffoli and Fredkin logic gates}.
\newblock {\em Proc. Natl. Acad. Sci.}, 109:21228--21233, 2012.

\bibitem{Liu2018}
Tong Liu, Bao-Qing Guo, Chang-Shui Yu, and Wei-Ning Zhang.
\newblock {One-step implementation of a hybrid Fredkin gate with quantum
  memories and single superconducting qubit in circuit QED and its
  applications}.
\newblock {\em Opt. Express}, 26:4498, 2018.

\bibitem{Siegman1986}
A.~E. Siegman.
\newblock {\em Lasers}.
\newblock University Science Books, 1986.

\bibitem{Durnin1987}
J.~Durnin, J.~J. Miceli, and J.~H. Eberly.
\newblock Diffraction-free beams.
\newblock {\em Phys. Rev. Lett.}, 58:1499--1501, 1987.

\bibitem{Gibson:04}
Graham Gibson, Johannes Courtial, Miles~J. Padgett, Mikhail Vasnetsov, Valeriy
  Pas'ko, Stephen~M. Barnett, and Sonja Franke-Arnold.
\newblock Free-space information transfer using light beams carrying orbital
  angular momentum.
\newblock {\em Opt. Express}, 12:5448--5456, 2004.

\bibitem{Wang2012}
Jian Wang, Jeng~Yuan Yang, Irfan~M. Fazal, Nisar Ahmed, Yan Yan, Hao Huang,
  Yongxiong Ren, Yang Yue, Samuel Dolinar, Moshe Tur, and Alan~E. Willner.
\newblock {Terabit free-space data transmission employing orbital angular
  momentum multiplexing}.
\newblock {\em Nat. Photonics}, 6:488--496, 2012.

\bibitem{Bozinovic2013}
Nenad Bozinovic, Yang Yue, Yongxiong Ren, Moshe Tur, Poul Kristensen, Hao
  Huang, Alan~E. Willner, and Siddharth Ramachandran.
\newblock Terabit-scale orbital angular momentum mode division multiplexing in
  fibers.
\newblock {\em Science}, 340:1545--1548, 2013.

\bibitem{Wang2016}
Jian Wang.
\newblock {Advances in communications using optical vortices}.
\newblock {\em Photonics Res.}, 4:B14, 2016.

\bibitem{Garcia-escartin2013}
Juan~Carlos Garcia-Escartin and Pedro Chamorro-Posada.
\newblock {SWAP test and Hong-Ou-Mandel effect are equivalent}.
\newblock {\em Phys. Rev. A}, 87:052330, 2013.

\bibitem{Lovitz2018}
Benjamin Lovitz and Norbert L{\"{u}}tkenhaus.
\newblock {Families of quantum fingerprinting protocols}.
\newblock {\em Phys. Rev. A}, 97:032340, 2018.

\bibitem{Cincio2018}
Lukasz Cincio, Yiğit Subaşı, Andrew~T Sornborger, and Patrick~J Coles.
\newblock {Learning the quantum algorithm for state overlap}.
\newblock {\em New J. Phys.}, 20:113022, 2018.

\bibitem{Mirhosseini2013}
Mohammad Mirhosseini, Omar~S. Maga{\~{n}}a-Loaiza, Changchen Chen, Brandon
  Rodenburg, Mehul Malik, and Robert~W. Boyd.
\newblock {Rapid generation of light beams carrying orbital angular momentum}.
\newblock {\em Optics Express}, 21:30196, 2013.

\bibitem{Allen1992}
L.~Allen, M.~Beijersbergen, R.~Spreeuw, and J.~Woerdman.
\newblock {Orbital angular momentum of light and the transformation of
  Laguerre-Gaussian laser modes}.
\newblock {\em Phys. Rev. A}, 45:8185--8189, 1992.

\bibitem{Torres2012}
Juan~P. Torres.
\newblock {Multiplexing twisted light}.
\newblock {\em Nat. Photonics}, 6:420--422, 2012.

\bibitem{Gonzalez2006}
N.~Gonz{\'{a}}lez, Gabriel Molina-Terriza, and Juan~P. Torres.
\newblock {How a Dove prism transforms the orbital angular momentum of a light
  beam}.
\newblock {\em Opt. Express}, 14:9093, 2006.

\bibitem{Fuchs1999}
Christopher~A Fuchs and Jeroen Van~De Graaf.
\newblock {Cryptographic Distinguishability Measures}.
\newblock {\em IEEE Trans. Inf. Theory}, 45:1216, 1999.

\bibitem{Turtaev2017}
Sergey Turtaev, Ivo~T. Leite, Kevin~J. Mitchell, Miles~J. Padgett, David~B.
  Phillips, and Tom{\'{a}}{\v{s}} {\v{C}}i{\v{z}}m{\'{a}}r.
\newblock {Comparison of nematic liquid-crystal and DMD based spatial light
  modulation in complex photonics}.
\newblock {\em Opt. Express}, 25:29874, 2017.

\bibitem{Jha2011}
Anand~Kumar Jha, Girish~S. Agarwal, and Robert~W. Boyd.
\newblock {Supersensitive measurement of angular displacements using entangled
  photons}.
\newblock {\em Phys. Rev. A}, 83:053829, 2011.

\bibitem{Slussarenko2010}
S.~Slussarenko, V.~D'Ambrosio, B.~Piccirillo, L.~Marrucci, and E.~Santamato.
\newblock {The Polarizing Sagnac Interferometer: a tool for light orbital
  angular momentum sorting and spin-orbit photon processing}.
\newblock {\em Opt. Express}, 18:27205, 2010.

\end{thebibliography}
 
\end{document}